\documentclass[review]{elsarticle}

\usepackage{lineno,hyperref}
\usepackage{underscore}
\usepackage{amssymb}
\usepackage{amsmath}
\usepackage{color}
\usepackage{epstopdf}
\usepackage{graphicx}

\journal{arXiv}

\begin{document}

\begin{frontmatter}

\title{Predicting Critical Nodes in Temporal Networks by Dynamic Graph Convolutional Networks}
\author{En-Yu Yu}
\address{Big Data Research Center, University of Electronic Science and Technology of China, Chengdu 611731, P. R. China}
\author{Yan Fu}
\address{Big Data Research Center, University of Electronic Science and Technology of China, Chengdu 611731, P. R. China}
\author{Jun-Lin Zhou}
\address{Big Data Research Center, University of Electronic Science and Technology of China, Chengdu 611731, P. R. China}
\author{Hong-Liang Sun}
\address{School of Information Engineering, Nanjing University of Finance and Economics, 210023, P. R. China}
\address{School of Computer Science and Technology, University of Nottingham, Ningbo, 315100, P. R. China}  
\author{Duan-Bing Chen*}
\address{Big Data Research Center, University of Electronic Science and Technology of China, Chengdu 611731, P. R. China}
\address{The Research Base of Digital Culture and Media, Sichuan Provincial Key Research Base of Social Science, Chengdu 611731, P. R. China}
\address{Union Big Data Tech. Inc., Chengdu 610041, P. R. China}
\address{Correspondence should be addressed to dbchen@uestc.edu.cn}

\begin{abstract}
Many real-world systems can be expressed in temporal networks with nodes playing far different roles in structure and function and edges representing the relationships between nodes. Identifying critical nodes can help us control the spread of public opinions or epidemics, predict leading figures in academia, conduct advertisements for various commodities, and so on. However, it is rather difficult to identify critical nodes because the network structure changes over time in temporal networks. In this paper, considering the sequence topological information of temporal networks, a novel and effective learning framework based on the combination of special GCNs and RNNs is proposed to identify nodes with the best spreading ability. The effectiveness of the approach is evaluated by weighted Susceptible-Infected-Recovered model. Experimental results on four real-world temporal networks demonstrate that the proposed method outperforms both traditional and deep learning benchmark methods in terms of the Kendall $\tau$ coefficient and top $k$ hit rate.
\end{abstract}


\end{frontmatter}

\section{Introduction}
Nowadays, people’s lives are closely related to various complex networks, such as social\cite{Weng2010}, traffic\cite{Ghosh2011} and email\cite{Guimera2003} networks. Network science is a vast and interdisciplinary research field which gradually become a hot topic in many branches of sciences. Rich-club\cite{Colizza2006detecting} shows that only a few critical nodes are needed to effectively affect and control the structure and function of the network. Therefore, to identify important nodes is thus significant, allowing us find influential spreaders\cite{Gallos2010}, control propagation of rumors\cite{Zhou2019} and plan precise marketing\cite{Zhang2019}. In recent years, many critical nodes identification methods in static networks are proposed\cite{lu2016vital, Guo2020, Chen2019}. Traditional methods mainly focus on network structure and information dissemination. Researcher are dedicated to find critical nodes by some heuristic algorithms, such as degree centrality\cite{PhillipBonacich1972}, betweenness centrality\cite{C1977} and k-shell\cite{Gallos2010}. With the development of deep learning, increasing number of researchers are beginning to solve problems in their own fields with the help of deep learning. Representation learning\cite{Cui2019} based methods in terms of embedding node as vectors or matrices, then design suitable learning frameworks to learn features of critical nodes, such as RCNN\cite{Yu2020}, InfGCN\cite{Zhao2020} and FINDER\cite{Fan2020}. All types of methods have good performance on various static networks. 

Most current methods for identifying critical nodes focus on static networks, yet most networks in the real world change over time. Temporal networks are not just an extension of static networks, which contain more in-depth and detailed information\cite{Holme2012, Michail2018}. However, in temporal networks, the identification of critical nodes is not a trivial task. We can easily get the critical nodes in the current network by static methods, but we have no way of knowing whether the current critical nodes are still important in the future. Vital nodes can be identified in static networks, but can only be predicted in temporal networks. At present, there are two main research topics in this area:(1) methods based on structure and propagation dynamics\cite{Pan2011}; (2) methods derived from dynamic graph neural networks (DGNNs)\cite{Skarding2020}.

In order to exploit both structured data and temporal information, a new learning framework named dynamic graph convolutional networks (DGCNs) is proposed. Our approach is based on the combination of graph convolutional networks (GCNs) and a Long Short-Term Memory networks (LSTM)\cite{Gers2003}. By combining the structural features learned by GCNs with the temporal features learned by LSTM, DGCNs can well predict the node which has stronger spreading ability in future. The performance of DGCNs is compared with node2vec\cite{Grover2016}, struc2vec\cite{Ribeiro2017}, temporal dynamics-sensitive centrality(TDC)\cite{Huang2017Predicting} and temporal k-shell(TK)\cite{Ye2017}, by weighted SIR model\cite{Kimura2009} on four real-world temporal networks. Experimental results suggest that DGCNs can effectively predict nodes with the best spreading ability and significantly outperforms benchmark methods in terms of Kendall $\tau$ coefficient and top $k$ hit rate. Moreover, the training time of DGCNs is linearly related to the size of networks and can be used for large networks.

The structure of this paper is as follows. Section 2 is a discussion on related works. Section 3 is the background for temporal networks, RNNs and GCNs. Section 4 is the detailed description for our method. Sections 5 is experimental results with analysis and discussion. Finally, conclusions are drawn in Section 6.

\section{Related Works}
To deal with the problems of identifying critical nodes in temporal networks, an intuitive idea is to extend methods in static networks such as degree, closeness, and betweenness centrality to temporal networks. By this idea, Kim et al.\cite{Kim2012}proposed the time-ordered graph, embedding dynamic networks into directed and static networks. And Huang et al\cite{Huang2017Predicting} proposed the temporal version of dynamic-sensitive centrality, which extends dynamic-sensitive centrality\cite{Liu2016} to temporal networks by the Markov chain for the epidemic model. By coupling centrality matrices in each snapshots into a supracentrality matrix, Taylor et al.\cite{Taylor2017} proposed an extension framework for static centrality measures such as eigenvector-based centrality. Huang et al.\cite{Huang2017} defined a supra-evolution matrix to describe the structure of temporal networks, which effectively reduces the computational complexity. This type of methods can find critical nodes in the current temporal network but cannot predict the importance of nodes in the future. 

In recent years, graph neural networks (GNNs) have become a new research hotspot and been used to solve graph-related problems, such link
prediction\cite{Ma2018Dynamic, Zhang2018Link}, graph\cite{Niepert2016, Kipf2017} and node\cite{Grover2016, Ribeiro2017} classification. Among them, DGNNs\cite{Kazemi2020} are recently prevailing deep learning models used to deal with temporal graphs, which often make use of a graph neural network (GNNs)\cite{Wu2021} and a recurrent neural network (RNNs)\cite{Medsker2013}. GCRN-M\cite{Seo2018} stacks a spectral GCN\cite{MichaelDefferrard2016} and a standard LSTM to predict structured sequences of data. DyGGNN\cite{Taheri2019} uses a gated graph neural network (GGNN)\cite{Li2015}combined with a standard LSTM to learn the evolution of dynamic graphs. Chen et al.\cite{Chen2018Graph} present GC-LSTM, which preforms a spectral GCNs on the hidden layer of the standard LSTM for dynamic link prediction. At present, DGNNs are mainly aimed at learning representations of entire dynamic graphs, but the learning framework specific to nodes in temporal networks is still lacking. 

\section{Background}

\subsection{Temporal Networks}
Temporal networks can be divided into continuous-time representation and discrete-time representation. In this paper, we only consider the discrete-time temporal networks. A discrete-time temporal network $TG \in [0, T]$ can be defined as a set of ordered static networks (snapshots). That is, $TG=\{G^1,G^2,...,G^L\}$ where $L=T/\delta$ represents the number of snapshots, which is determined by the time span $T$ and the time interval $\delta$ of each snapshot.
$G^t=\{V^t,E^t\}$ represents the spanning subgraph which consists of nodes and edges appearing in $[\delta \cdot (t-1), \delta \cdot t]$. 

\subsection{Recurrent Neural Networks}
Recurrent neural networks (RNNs)\cite{Schmidhuber2015} are a type of neural networks with a hidden layer that recur over time and often used to process sequence data such as the stock data. Unlike other neural networks, RNNs implement the structure which retains a certain memory of past information. Among of all RNNs, Bidirectional RNN (Bi-RNN)\cite{Schuster1997} and Long Short-Term Memory networks (LSTM)\cite{Hochreiter1997} are widely used. The standard RNNs is based on a simple repeating cell. LSTM extend the repeating cell by combining four interacting units.

\subsection{Graph Neural Networks}
Graph neural networks (GNNs)\cite{Zhang2018} are to combine graph data with neural networks, and perform calculations on graph data. Graph convolutional networks (GCNs) are one of GNNs which root in convolutional neural networks (CNNs)\cite{LeCun1998} and usually used to extract all levels of graph representation and perform graph classification tasks. GCNs follow the framework of exchanging information with neighbors and the key to it is how to aggregate the node features from its neighborhood.

\section{Method}
\subsection{Problem Definition}
The problem of predicting important nodes can be converted to a regression problem in deep learning. Suppose temporal networks $TG=\{G^1,G^2,...,G^L\}$, a function $f$ need to learn:
 \begin{equation}
Score_t = f(G^{t-s},G^{t-s+1},...,G^{t-1}) , t \in [s+1, L+1]
\end{equation}
where $Score_t$ is the predicted scores for nodes which have appeared before time $t$ in $G^t$ and $s$ is the number of input snapshots. We konw that GCNs can effectively deal with static graph data and RNNs is good at handling sequence data.
If the importance of a node in each snapshot can be learned through GCNs, then we can get a sequence containing the importance of the node in each snapshot. Finally, the importance of the node in the future can be predicted by using the sequence as the input of RNNs. Inspired by the idea of combining an special of GCNs and RNNs, a new learning framework named dynamic graph convolutional networks (DGCNs) is proposed to predict critical nodes in temporal networks.

\subsection{Dynamic Graph Convolutional Networks}
In this subsection, a framework DGCNs for learning representations from arbitrary temporal networks is proposed. The process of DGCNs is shown in Fig \ref{fig:DGCNs}. Step weighted snapshots to step CNNs layer can be regarded as a special GCNs which implicitly outputs the importance of nodes at time $t$, essentially a process of node embedding. And the details are as follows :

 \begin{figure}[ht]
	\centering
	\includegraphics[width=12cm, height=7cm]{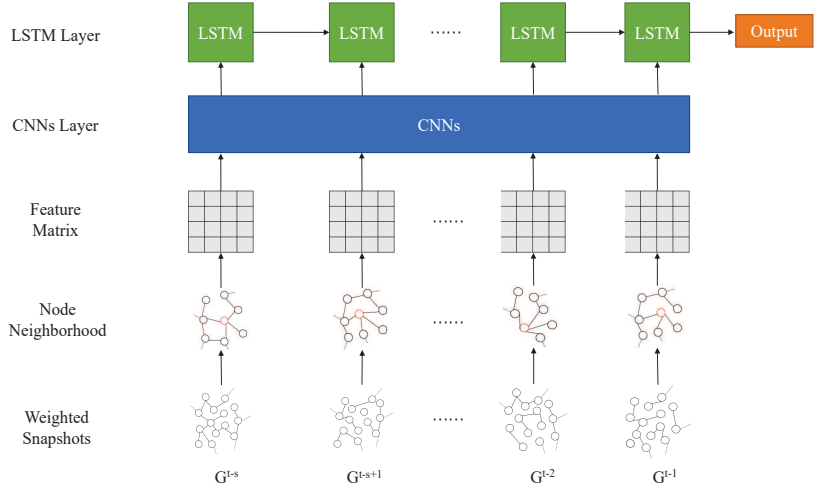}
	\caption{An illustration of DGCNs}
	\label{fig:DGCNs}
\end{figure}
\textbf{Weighted Snapshots}: Divide the temporal network $TG$ into $L$ snapshots according to the time interval $\delta$, $TG=\{G^1,G^2,...,G^L\}$. In the standard snapshot $G^t=\{V^t,E^t\}$, $ E^t$ contains all edges that appear in the time interval $[\delta \cdot (t-1), \delta \cdot t]$. This processing method is coarse-grained, because there may be multiple contacts between nodes in a time interval, but it is only retained once in the snapshot. In order to more accurately describe the relationship between a node and its neighbors in a temporal network, weighted snapshots are proposed. In a weighted snapshot $G^t=\{V^t,E^t,W^t \}$, $W^t$ records the number of occurrences of each edge in the time interval $[\delta \cdot (t-1), \delta \cdot t]$. $W^t(u, v) = w $ only if  $e_{u, v}$ appears $w$ times in $[\delta \cdot (t-1), \delta \cdot t]$. In addition, in order to ensure data consistency, each snapshot contains all nodes in this paper.

\textbf{Node Neighborhood}: Suppose the size of a neighborhood is $k$, that is, we need find $k-1$ neighbors for each node and number neighbors according to order. The strategy for selecting neighbors from a snapshot: the first priority is the distance between the node and the neighbor. In the same distance, prioritize neighbors with high degree. After selecting neighbors, for each node, such as node $u$, generate the subnetwork $G^t_u$ which contains node $u$ and its $k-1$ neighbors from the snapshot $G^t$.

\textbf{Node Feature Matrix}: For each subnetwork, such as $G^t_u$,  $\textbf{A}^t_{u}$ is the adjacency matrix of $G^t_u$ and the feature matrix of node $u$ is defined as:
 \begin{equation}
\textbf{B}^t_{u}[i, j]=\left\{\begin{matrix}
O^t_{u_i}, & i=j=1,2,...,k\\ 
\textbf{A}^t_{u}[i,j] \cdot W^t(u_i,u_j), & other \quad cases
\end{matrix}\right.
\end{equation}
where $u_1, u_2, . . . , u_{k}$ are node $u$ and its $k-1$ neighbors respectively. $O^t_{u_i}$ is the out degree of node $u_i$ in the snapshot $G^t$. In addition, if  $G^t$ is disconnected, we might not find enough neighbors for node $u$, and we need expand $\textbf{B}^t_{u}$ with zero padding.

\textbf{CNNs Layer}: The CNNs in this layer is same as the CNNs in $RCNN$\cite{Yu2020}. Through the convolution processing of feature matrices, the importance of nodes can be implicitly expressed. There are two reasons why we use the same CNNs to train nodes in different snapshots instead of training a separate CNNs for each snapshots: (1) Global training does not change the importance ranking of nodes in the same snapshot, and can effectively reduce training parameters. (2) By increasing the number of input snapshots $s$, a large number of training nodes can be quickly obtained.

\textbf{LSTM Layer}: A standard LSTM\cite{hochreiter1997long} is used in this layer. And the input element in time $t$ is $x_t = CNN(\textbf{B}^t_{u})$. In addition, the input size is 1, hidden size is 64, number of recurrent layers is 2. Finally, a fully connected layer is used to output the predicted score for nodes.

\textbf{Label}: In previous researches, the influence of nodes over a period of time is usually measured by SIR spreading model. In this paper, we use weighted SIR model to measure the influence of nodes in discrete-time temporal networks. In SIR model, nodes have three states, i.e., Susceptible, Infected and Recovered. In time interval $[\delta \cdot (t-1), \delta \cdot t]$,  
infected nodes will infect their susceptible neighbors with a probability $\beta$ and recover with a probability $\mu$.  In weighted SIR model, the susceptible node $u$ will be infected by the infected neighbor $v$ with $1-(1-\beta)^{W^t(v, u)}$. Then, $v$ will recover with $\mu$. $N^t_u(x)$ is defined as the number of recovered and infected nodes after $x$ intervals under weighted SIR model from the initial infected node $u$ in snapshot $G^t$. We can use $N^t_u(x)$ to represent the importance of  $u$ in snapshot $G^t$.

Suppose we have a temporal networks $TG=\{G^1,G^2,...,G^L\}$ and want to predict the importance of nodes in snapshot $G^t$( $t\leq L-10$ when use $G^t$ as the testing set). The data that can be used for training are  $\{G^{1},G^{2},...,G^{t-1}\}$.
Because we need to use 10 snapshots to generate the label, we can use $\{G^{t-11-s},...,G^{t-11}\}$ as the training set and $N^{t-10}(10)$ as the label. And the number of input snapshots $s$ is also need to train by traversing $L$. The loss function is squared loss function.

\subsection{Complexity Analysis}
Let the temporal networks be $TG=\{G^1,G^2,...,G^L\}$, the size of the neighborhood be $k (k<<n)$ and use $s$ snapshots to train DGCNs. For each snapshot and node, the time complexity of generating the feature matrix is $O(k^2)$, the time complexity of CNNs is $O(k^2)$\cite{Yu2020} and the time complexity of LSTM is $O(n)$. So the time complexity of DGCNs is $O(\sum_{s=1}^{s=L}s \cdot I\cdot k^2\cdot N )=O(L^2 \cdot I\cdot k^2\cdot N)$, where $I$ is the number of training iterations. Actually, when all parameters are fixed, The complexity of DGCNs is linearly related to the number of nodes in the temporal network.

\section{Experiments}
In this section, datasets, experimental settings and results achieved by DGCNs are described in detail. 
\subsection{Datasets}
In our experiments, we selected four real-world temporal networks as public datasets. (1)Email\cite{Paranjape2017}. The directed temporal network is generated by the mail data from a research institution in European. (2)Contact\cite{Chaintreau2007}. An undirected temporal network contains connections between users using mobile wireless devices. an edge is generated when two people are in contact. (3)DNC\cite{Yu2020SR}. This is a directed temporal network of emails in the 2016 Democratic National Committee email leak. (4) UCI\cite{Opsahl2009}. An directed temporal network contains sent messages between students at the University of California, Irvine. Some basic features of these networks are listed in table \ref{tb.data}.

\begin{table}
	\centering
	\caption{Statistical properties of four real-world networks. $N$ and $M$ are the total number of nodes and edges, respectively. $\delta$ is the time interval(hours) used to divide networks and $L$ is the number of snapshots}\label{tb.data}
	\begin{tabular}{l llll ll}
		\hline
		Networks&$N$&$M$&$\delta $&$L$&\\ \hline
		Email      & 986      &  332334     &    168     & 75     \\
		Contact     & 274      &  28244     &  1      & 69   \\
		DNC      & 1865      & 39623     & 12       & 56     \\
		UCI     & 1899      &  59835     &  24       & 58      \\
		\hline
	\end{tabular}
\end{table}

\subsection{Experimental Settings}
In this paper, for each dataset, training set is $\{G^{31-s},G^{31-s+1},...,G^{30}\}, s \in [1, 30]$ with label $N^{31}(10)$(training infection rate $\beta_t$=0.05) and testing set is $\{G^{41-s},G^{41-s+1},...,G^{40}\}$ with label $N^{41}(10)$. That is, for DGCNs and all benchmark methods, the snapshots $G^t, t > 40$ is unknown, only $G^t, t \leq 40$ can be used to predict the importance of nodes in $G^{41}$.  Finally, compare the prediction results with $N^{41}(10)$ to verify the performance of methods. In addition, the GPU for all experiments is 1600 MHz with 8G memory.

\subsection{Benchmark Methods}
In order to effectively evaluate the performance of DGCNs , benchmark methods include two types, one is based on network structure and propagation dynamics, and the other is based on graph neural networks.

The temporal k-shell(TK)\cite{Ye2017} is defined as
 \begin{equation}
TK(v)=  \sum_{u\in \Gamma_{v}}\sum_{t=1}^{L}min\left \{k _{s}^{t}(v), k_{s}^{t}(u)\right \}
\end{equation}
where $k_{s}^{t}(v)$ is the k-shell of node $v$ and $\Gamma_{v}$ is the neighbors of node $v$ in snapshot $t$.

Temporal dynamics-sensitive centrality(TDC)\cite{Huang2017Predicting} is defined as
\begin{equation}
S =\sum_{r=0}^{L-1}\beta H_{*}^{r}\textbf{A}^{r+1}X
\end{equation}

\begin{equation}
H_{*}^{t}=\prod_{\alpha =t}^{1}[\beta \textbf{A}^{\alpha}+(1-\mu )I], H_{*}^{0} = 1
\end{equation}
where $S_i$ is the score of node $i$ and $X=(1,1,...1)^{T}$.

The core of DGCNs is how to embed nodes in snapshots, and can be replaced by other well-known node embedding methods, such as node2vec\cite{Grover2016} and struc2vec\cite{Ribeiro2017}. In this paper, the replaced methods are called N2V-LSTM and S2V-LSTM. These two methods embed nodes of each snapshot and input them into LSTM. 


\subsection{Results}
As we know, for the problem of predicting the importance of nodes in temporal networks, the closer the predicted ranking is to the real ranking, the better the performance of the method is. Here we use the kendall $\tau$ correlation coefficients\cite{Knight1966} to measure the performance of all methods. 

\begin{figure}[ht]
	\centering
	\includegraphics[width=6cm, height=6cm]{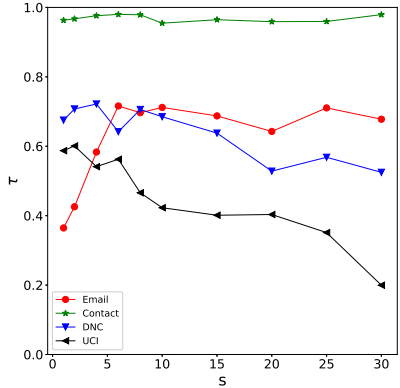}
	\caption{The impact of the number of input snapshots $s$ on final kendall $\tau$.}
	\label{fig:s}
\end{figure}

Firstly, we perform a sensitivity analysis of the number of input snapshots $s$ on DGCNs. In Fig \ref{fig:s}, we show the impact of $s$ on final kendall $\tau$ between the real ranking and predicted ranking with the infection rate $\beta=0.05$. From Fig \ref{fig:s}, in Email network, the time interval is 168 hours(a week), when $s$ is less than 5, the performance of DGCNs improves as $s$ increases, and then it stays in a steady state. This means that the importance of a user in the next 10 weeks can be predicted only by using the data in the past 5 weeks and more data will not improve the performance. In Contact network, the time interval is 1 hours, DGCNs is almost unaffected by $s$ while performing extremely well. This also means that we only need the data of past 1 hour to predict the importance of the user in next 10 hours very accurately. In DNC network, the time interval is 12 hours,  it can be seen that the performance of DGCNs reaches its peak at $s=4$ and then shows a downward trend. This explains we can use the data of past 48 hours to predict the importance of the user in next 120 hours. However, using too old historical data will bring a lot of noise, which will seriously affect the performance of DGCNs. Similar to the DNC network, in UCI network, the time interval is 24 hours(1 day),  the performance of DGCNs reaches its peak at $s=2$ and then shows a downward trend. These results show that in different temporal networks, the historical data needed to predict the importance of nodes is not as much as possible. The data which is too old may even have a negative impact on the prediction results. 

\begin{figure}[ht]
	\centering
	\includegraphics[width=12cm, height=12cm]{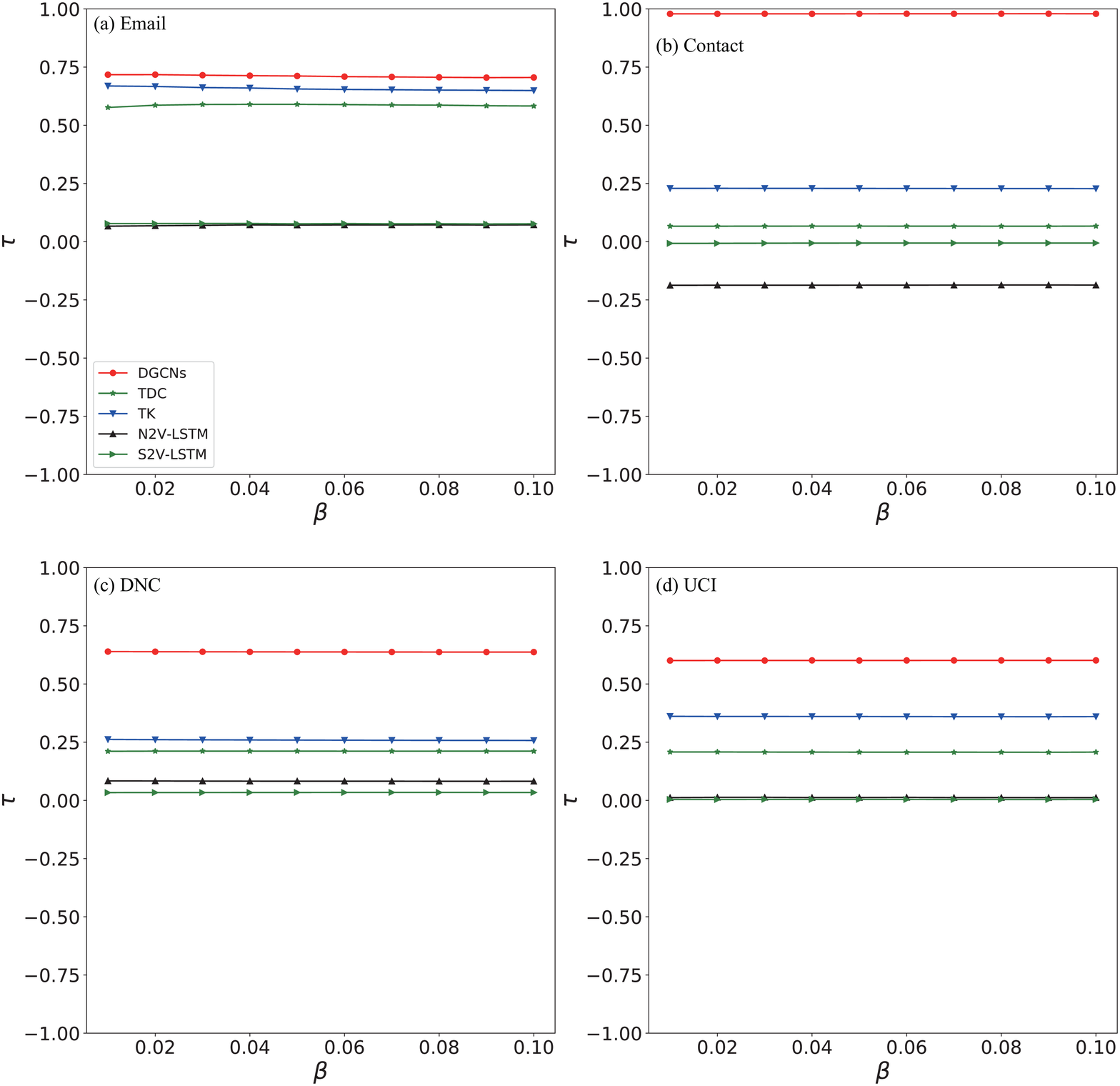}
	\caption{The impact of infection rate $\beta$ on final kendall $\tau$.}
	\label{fig:corr}
\end{figure}

After the training of $s$ is completed, from Fig \ref{fig:corr}, compared with other methods, no matter how the infection rate $\beta$ changes, DGCNs has the highest kendall $\tau$ correlation coefficients. This means that among all the methods, the ranking result of DGCNs is the closest to the real ranking result. In addition, we can see that the performance of N2V-LSTM and S2V-LSTM is very poor in all networks, especially in Contact. This shows that a simple combination of deep learning models cannot solve some professional problems in the field of complex networks. An effective and feasible learning framework needs to be designed with more knowledge in related fields.
 
\begin{figure}[ht]
	\centering
	\includegraphics[width=12cm, height=12cm]{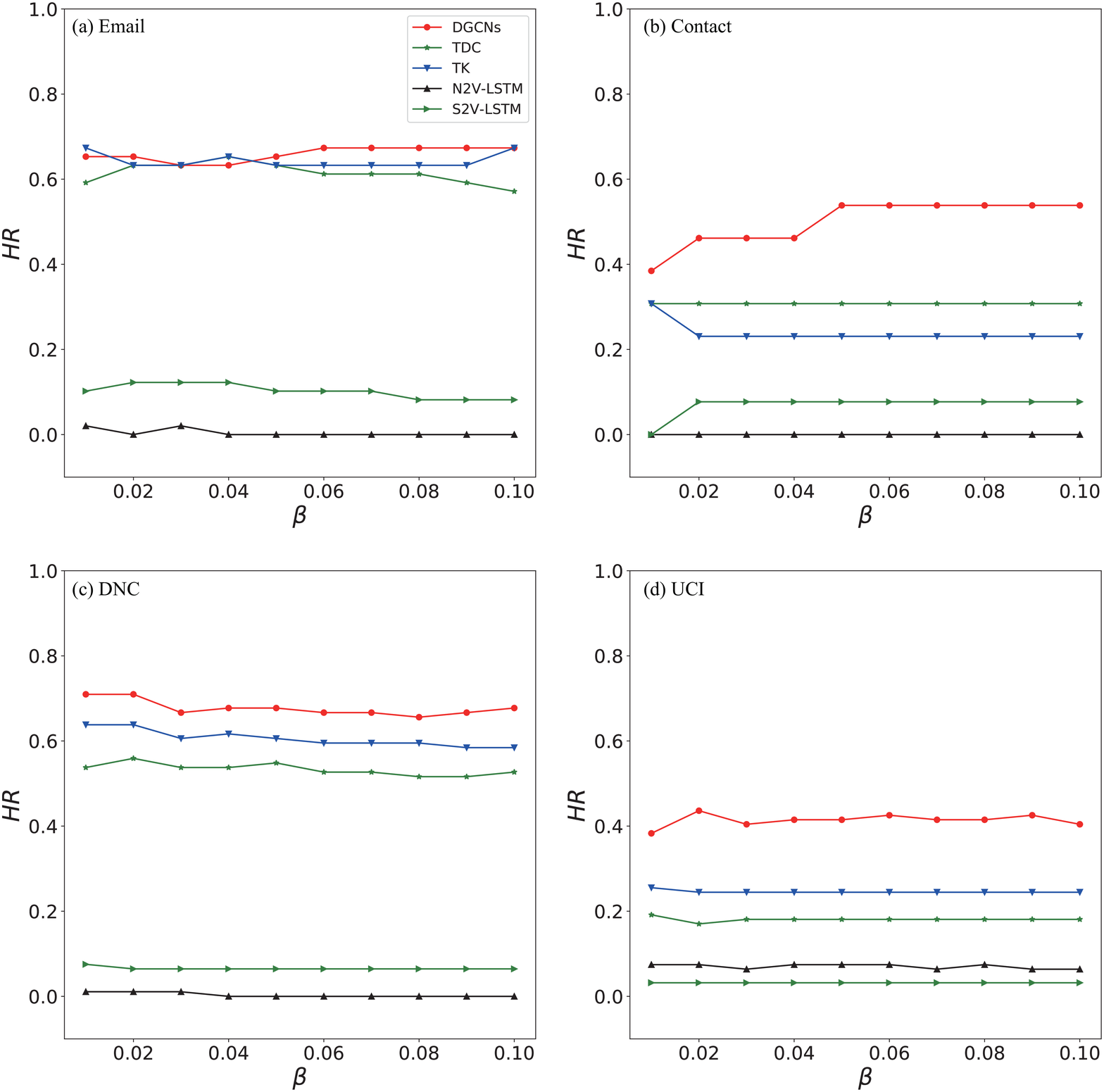}
	\caption{ The top5\% $HR$ versus different infection rate $\beta$.}
	\label{fig:HR}
\end{figure}

In real life, people usually care more about top $k$ nodes than the ranking of nodes. So we compared the hit rate of top $k$ nodes of all methods. For details, The evaluation index is the top $k$ hit rate and defined as
\begin{equation}
HR =\frac{|P\bigcap R|}{|R|}
\end{equation}
where $P$ is the set of top $k$ nodes in predicted ranking and $R$ is the set of top $k$ nodes in real ranking. Obviously, the best method should have the largest $HR$. From Fig \ref{fig:HR}, it can be seen that DGCNs outperforms other methods in most cases under different infection rate $\beta$. These results further proves the effectiveness of DGCNs.

Finally, The time cost of the above five methods is compared. Table \ref{tb.trainTime} shows the training time of $DGCNs$ on four temporal networks with different $s$. And it can be seen that DGCNs spends very little training time. And Table \ref{tb.testTime} shows the time cost of ranking nodes by all methods. It's not hard to see $RCNN$ achieves the best predicting results in a reasonable time.

\begin{table}
	\centering
		\caption{The training time(senconds) of four temporal networks with different $s$.}\label{tb.trainTime}
	\begin{tabular}{l llll ll}
			\hline
		Networks & $s=1$ & $s=2$ & $s=4$ & $s=8$ & $s=15$ & $s=30$ \\ \hline
		Email     & 69.17      &  77.45     &  94.46 &  132.30     & 195.10    & 341.90    \\
		Contact       & 21.54      & 22.73     & 28.06 &  38.87      & 58.55    &  106.61    \\
		DNC     & 128.01      &  147.25    &  175.50   &  247.95    & 362.78    &  643.84    \\
		UCI       & 126.32      &  142.78    & 176.30   & 261.63    & 385.58    &  752.10   \\
		\hline
	\end{tabular}
\end{table}

\begin{table}
	\centering
		\caption{The time cost of ranking nodes (senconds) on different networks by five methods.}\label{tb.testTime}
	\begin{tabular}{l llll ll}
		\hline
		Networks&$DGCNs$&N2V-LSTM&S2V-LSTM&TDC&TK \\ \hline
		Email     & 2.46      &  0.08     &  0.07      & 26.67    & 0.32    \\
		Contact      & 0.09      &0.04     & 0.04       & 1.20    &  0.06    \\
		DNC     & 0.45      &  0.16    &  0.14       & 132.71    &  0.37    \\
		UCI      & 0.23      &  0.17     &  0.13      & 140.74    &  0.36    \\
		\hline
	\end{tabular}
\end{table}

\section{Conclusions}
In this work, we introduce a new learning framework DGCNs that can predict the importance of nodes in temporal networks. The model consists of  a special graph convolutional networks and long-short term memory networks. We have assessed the performance of DGCNs on four real-world networks against some benchmark methods. The results show that the ranking by DGCNs is the closest to the real ranking and has the highest top $k$ hit rate. What is more, the training time complexity of DGCNs is linearly related to the number of nodes and can be used for large-scale networks. Like most current models,  DGCNs rely on snapshots, which are actually a relatively crude temporal representations. Continuous time methods can often capture more in-depth and detailed information. So, how to extend DGCNs to continuous-time temporal networks is one of the main tasks in the future. What's more, another interesting extensions of our work may consists in critical edges in temporal networks.

\clearpage

\section*{Acknowledgements}
This work is jointly supported by the National Natural Science Foundation of China under Grant Nos. 61673085 and 71901115, by the Science Strength Promotion Programme of UESTC under Grant No. Y03111023901014006, by the International Cooperation Programme of JiangSu Province under Grant No BZ2020008,  by the Young Scholar Programme from NUFE under Grant No SHLXW19001 and by the National Key R\&D Program of China under Grant No. 2017YFC1601005.

\bibliography{mybibfile}

\end{document}